\documentstyle[amssymb,a4,12pt]{article}
\begin{document}
\renewcommand{\theequation}{\thesection.\arabic{equation}}
\begin{titlepage}
\noindent
{\Large\bf A Gallavotti-Cohen Type Symmetry in the Large}
\begin{center}
{\Large\bf	Deviation Functional
for Stochastic Dynamics}
\bigskip\bigskip\bigskip\\
{\large Joel L. Lebowitz}\bigskip\\
Department of Mathematics and Physics, Rutgers University,\\
Hill Center, New Brunswick, New Jersey 08903, USA\\email:
lebowitz@math.rutgers.edu\medskip\bigskip\\
{\large Herbert Spohn}\bigskip\\
Zentrum Mathematik, TU M\"{u}nchen,\\
D-80290 M\"{u}nchen, Germany\\
email: spohn@mathematik.tu-muenchen.de
\end{center}


\vspace*{4cm}
\noindent
{\bf Abstract.} 
We extend the
work of Kurchan on the Gallavotti-Cohen fluctuation theorem, which yields a
symmetry property of the large deviation function, to general Markov
processes.  These include jump processes describing the evolution of
stochastic lattice gases driven in the bulk or through particle reservoirs,
general diffusive processes in physical and/or velocity space, as well as
Hamiltonian systems with stochastic boundary conditions.  For dynamics
satisfying local detailed balance we establish a link between the action
functional of the fluctuation theorem and the entropy production.  This
gives, in the linear regime, an alternative derivation of the Green-Kubo
formula and the Onsager reciprocity relations.  In the nonlinear regime
consequences of the new symmetry are harder to come by and the large
derivation functional difficult to compute.  For the asymmetric simple
exclusion process the latter is determined explicitly
using the Bethe ansatz in the limit of large $N$.  
\end{titlepage}

\section{Introduction}

The study of stationary nonequilibrium states (SNS) of macroscopic systems
evolving according to classical (or even quantum) mechanics, which are kept
out of equilibrium by contact with thermal reservoirs has a long history.
The interactions of the system with the reservoirs are usually modeled by
the addition of stochastic boundary terms to the deterministic evolution
equations describing the isolated system \cite{BL}--\cite{GI}. Fully
deterministic evolutions of systems coupled to infinitely extended
reservoirs have also been investigated \cite{LS}--\cite{EPR}, as have fully
stochastic models \cite{KLS}--\cite{ELS}.  A more recent development has
been the study of SNS of systems evolving via deterministic
``thermostatted'' dynamics \cite{H,EM,CELS}. In this scheme the SNS is
maintained by deterministic external driving forces which are balanced by a
``friction term'' suitably chosen so that the system evolves on a compact
surface (generally one of constant energy) in phase space.  The resulting
dynamics no longer conserves phase space volume, which is contracted on the
average \cite{DR}.  Consequently, the SNS is singular with respect to the
Liouville (volume) measure induced on the surface.

While the jury is still out on what has or can be achieved by such a
dynamical systems approach, there are some intriguing theoretical results
which have emerged \cite{G,R}.  In particular Gallavotti and Cohen
\cite{GC} motivated by results of computer simulations \cite{ECM}
discovered that under suitable assumptions these SNS satisfy a certain
symmetry, which they doobed a ``fluctuation theorem''.  Assuming that the
dynamics satisfies {\it time reversal invariance} and is sufficiently
chaotic, so that the SNS is given by an SRB measure, they prove that the
probability distribution for the phase space contraction averaged along a
trajectory over the time span $\tau$ has, for large $\tau$, a highly
non-obvious symmetry, whose specific form will be given below. Near
equilibrium the fluctuation theorem implies Onsager reciprocity and the
Einstein relations. We refer to \cite{G1} for an exhaustive discussion of
the theorem and of the ``chaotic hypothesis'', which implies the validity
of the theorem for more general SNS.

In a recent work Kurchan showed that, with proper definitions, the
fluctuation theorem is valid also for certain diffusion processes
\cite{JK}.  In the present work we extend Kurchan's results and show that
for Markov processes the fluctuation theorem holds in great generality.
Our proof is based on the Perron-Frobenius theorem and goes much beyond the
cases considered in \cite{JK}.  While the assumed stochasticity of the
dynamics should provide ``for free'' a suitable modification of the chaotic
hypothesis it is far from obvious how to find the quantity that plays the
role of the phase space contraction.  In analogy to the thermostatted case
it is reasonable to expect that it will have some relation to the
production of Gibbs entropy.  In fact, this comes out from our general
construction under the additional assumption of local detailed balance.
This condition seems to play, in stochastic dynamics, the same role as time
reversal invariance in deterministic thermostatted dynamics.  When local
detailed balance is satisfied, then the action functional is proportional
to the current of the conserved quantity, e.g. in the case of bulk driven
lattice gases it is the bulk particle current. If local detailed balance is
not satisfied, then the action functional can still be defined but it does
not allow for such a direct physical interpretation.

Near equilibrium the fluctuation theorem, generalized to systems with
several currents, yields the Onsager symmetry and the usual Kubo formulae
for the linear transport coefficients.  These give a relation between
linear response and current fluctuations in equilibrium. In this sense the
fluctuation theorem, which holds also far from equilibrium, can be thought
of as a natural extension, in the spirit of GC, of the
fluctuation-dissipation theorem which holds for systems close to (local)
equilibrium.  In fact, when the large deviation functional is
approximately quadratic over a sufficiently large range of the driving
field, the fluctuation theorem does indeed yield such an approximate
extension.  We do not, however, have a quantitative criteria of when this
approximation is valid.  Even aside from this extension it would be of
great interest to have an experimentally verifiable consequence of the
novel symmetry predicted by the fluctuation theorem for the distribution of
large deviations in current carrying systems.  This seems difficult to
obtain for macroscopic systems since the time required for observing a
large deviation in such a system far exceeds any observation time.

The physical meaning of the fluctuation theorem can be understood best
through the application to simple models.  We therefore determine here the
action functional for which such a fluctuation theorem holds in various
examples: bulk and/or boundary driven lattice gases, the Fokker-Planck
equation with driving forces and a spatially varying temperature, and
Hamiltonian particle systems driven by boundary reservoirs.  In general,
the rate function appearing in the fluctuation theorem cannot be computed
explicitly and one has to rely on numerical simulations \cite{ECM,BGG}. In
stochastic models the task is easier as long as we stick to noninteracting
particles. We discuss one example of interacting particles, where the rate
function can be computed via the Bethe ansatz. When the size of the system
goes to infinity the rate function becomes singular.  This is analogous to
the behavior of thermodynamic potentials at a phase transition.  This is
done in the Appendix which may be read independently of the rest of the
paper.

Our analysis extends directly to discrete time stochastic processes, i.e.\
to probabilistic cellular automata.  In this situation it is natural to
think of the GC fluctuation theorem as a symmetry of discrete space--time
Gibbs measures, as has been pointed out by C. Maes \cite{M}.  In fact, the
framework of space--time Gibbs measures contains as special cases both the
stochastic dynamics studied here and the thermostatted systems satisfying
the chaotic hypothesis. In the latter case the space--time Gibbs measure is
constructed through a Markov partition of phase space.

\section{The fluctuation theorem for jump processes}
\setcounter{equation}{0}

We start with the fluctuation theorem for a stochastic time evolution
governed by a master equation on a state space with a finite number of
points. This minimizes technical complications and at the same time
provides a blue-print for stochastic systems with Langevin type
dynamics. In this section we investigate the abstract structure, specific
examples and applications will be discussed in later sections.\bigskip\\
{\large {\bf 2.1 Action functional, large deviations}} \medskip\\ We
consider a continuous time Markov jump process with finite state space
${\cal S}$. Points in ${\cal S}$ are denoted by $\sigma$. The jump process
is determined by the rates, $k(\sigma,\sigma') \ge 0$, for jumping from
$\sigma$ to $\sigma'$. More precisely, if the system is in the state
$\sigma$ it waits a random time $t \ge 0$ distributed according to the
exponential law $r(\sigma)e^{-r(\sigma)t} dt$, $r(\sigma) = \sum_{\sigma'}
k(\sigma,\sigma')$, and then jumps to $\sigma'$ with probability
$r(\sigma)^{-1}k(\sigma,\sigma')$, etc.. The generator acting on functions
$f:{\cal S} \rightarrow \Bbb{R}$ is given by
\begin{equation}
L f(\sigma) = \sum_{\sigma'} k(\sigma,\sigma') [f(\sigma') - f(\sigma)] 
\, .
\label{2.1}
\end{equation}
This means, if $\mu(\sigma,t)$ is the probability distribution
of $\sigma$ at time $t$,
then the rate of change of the average of $f$,
$\langle f \rangle_{\mu(t)} =
\sum_\sigma \mu(\sigma,t) f(\sigma)$, is given by
\begin{equation}
{d \over dt} \langle f \rangle_{\mu(t)} =
\langle Lf \rangle_{\mu(t)}\, .
\label{2.2}
\end{equation}
Taking $f$ to be a Kronecker delta at $\sigma$ (2.2) corresponds to
the master equation
\begin{equation}
{\partial \mu(\sigma,t) \over \partial t}  = \sum_{\sigma^\prime}
k(\sigma^\prime,\sigma) \mu(\sigma^\prime,t) - r(\sigma) \mu(\sigma,t)
= L^* \mu(\sigma,t)
\label{2.3}
\end{equation}
for the evolution of $\mu(\sigma,t)$, where $L^*$ denotes the adjoint of $L$.

We assume that if $k(\sigma,\sigma') > 0$, then also $k(\sigma',\sigma) >
0$, and that from every $\sigma$ all other $\sigma' \in {\cal S}$ can be
reached by a succession of steps with nonzero rates.  It follows then from
the general theory of finite state Markov processes that there exists a
unique stationary measure, $\mu_{{s}}$, which satisfies $ L^{*}
\mu_{s}(\sigma) = 0$ and is strictly positive, $\mu_{s} (\sigma) >0$.
Starting from any initial state $\sigma_0$, $\mu_s$ is approached
exponentially fast in time.

We could as well take the state space ${\cal S}$ to be countable or replace
${\cal S}$ by $\Bbb{R}$$^{n}$. In the latter case one prescribes the rates
$k(x,x')dx'$ for a jump from $x$ to the volume element $x' + dx'$.  Setting
$r(x) = \int k(x,x') dx'$,  the waiting time at $x$ is exponential with
rate $r(x)$ and the jump probability to $x'+dx'$ is $r(x)^{-1}
k(x,x')dx'$. Formally the difference is only in notation.  However, now one
has to make sure that the jump process has a unique invariant measure.

Let $s \rightarrow \sigma_{s}$ be a trajectory (or history)
of the jump process and define for it the quantity
\begin{equation}
        W(t,\{\sigma_s, 0\leq s\leq t\}) = \int_{0}^t
        \sum_{\sigma,\sigma'} w_{\sigma,\sigma'}(s) ds\, 
\label{2.4}
\end{equation}
with $w_{\sigma,\sigma'}(s)$  a sequence
of $\delta$-functions, located exactly
at those times $s$ when $\sigma_{s}$ jumps
from $\sigma$ to $\sigma'$,  with weight
\begin{equation}
  w(\sigma,\sigma') =   \log k(\sigma,\sigma') - \log 
  k(\sigma',\sigma)\, .
\label{2.5}
\end{equation}
This means, if the trajectory $\sigma_{s}$, $0\leq s \leq t,$ visits
in succession the states
$\sigma_0,\sigma_1,...,\sigma_n$,
where $\sigma_{0}$ is the state at time $0$ and $\sigma_n$
the one at time $t$, then
\begin{equation}
W(t,\{\sigma_s, 0\leq s\leq t\}) = \log
\left[ {k(\sigma_0,\sigma_1) \over k(\sigma_1,\sigma_0)} \dots
{k(\sigma_{n-1},\sigma_n) \over k(\sigma_n, \sigma_{n-1})}\right ] \, .
\label{2.6}
\end{equation}
(Note that at an allowed transition $k(\sigma,\sigma') > 0$ and,
by assumption, also\\
$k(\sigma',\sigma) > 0$).

For lack of a better name we call $W(t)$, and the similar quantities to be
defined below, an {\it action functional}.  (We will generally not indicate
explicitly the dependence of $W(t)$ on the stochastic trajectory
$\{\sigma_s,0\leq s \leq t\}$).

Let $\langle \cdot\rangle$ denote the expectation over all trajectories in
the stationary process, i.e.  starting in the steady state $\mu_{s}
(\sigma)$. We consider the generating function $\langle \exp [-\lambda
W(t)] \rangle$ of $W(t)$ and define
\begin{equation}
        \lim_{t \rightarrow \infty} - {1 \over t} \log \langle e^{-\lambda
W(t)} \rangle 
        = e(\lambda)\, .
\label{2.7}
\end{equation}
The {\it fluctuation theorem} we shall prove states that the limit (2.7)
exists with $e(\lambda)$  convex downwards,  and that
\begin{equation}
        e(\lambda) = e(1-\lambda)\, .
\label{2.8}
\end{equation}

By a general result, analogous to the equivalence of
ensembles in equilibrium statistical mechanics,
(2.7) implies a large deviation property for the probability
distribution $p_{t}(w)$ of $W(t)/t$, i.e. for large $t$ 
\begin{equation}
        p_{t}(w) \cong e^{-t \hat{e}(w)}\, ,
\label{2.11}
\end{equation}
where $\hat{e}$ is the Legendre transform of $e$. $\hat{e}(w)$ is convex
up, $\hat{e}(w) \geq 0$,
$\hat{e}(w_{0}) = 0$ for $w_{0} = lim_{t \to \infty} \langle {1 \over t}
W(t)\rangle$, 
and by (2.8) $\hat{e}$  satisfies
\begin{equation}
\hat{e}(w) - \hat{e}(-w) = -w\, ,
\label{2.12}
\end{equation}
i.e. the odd part of $\hat{e}$ is linear with slope $- 1/2$.

To be more precise, from (2.7) one concludes that for every interval
$I$
\begin{equation}
        \lim_{t \rightarrow \infty} - {1 \over t} \log {\rm Prob}
         ({1 \over t} W(t) \in I) =
        \min_{w \in I} \hat{e}(w)
\label{2.13}
\end{equation}
with
\begin{eqnarray}
        \hat{e}(w) & = & \max_{\lambda} \{e(\lambda) - \lambda w\} \nonumber\\
        & = & \max_{\lambda} \{e(1 - \lambda) - \lambda w\} = \max_{\lambda}
        \{e(\lambda) - (1-\lambda) w\} \nonumber \\
        & = & \hat{e} (-w) - w\, .
\label{2.14}
\end{eqnarray}

Eq. (2.10) is of the same form as the GC fluctuation theorem, with the
phase space contraction integrated along an orbit of the thermostatted
dynamics in GC replaced by (2.5) summed over the jumps along the stochastic
trajectory.  We note that the convexity of $e(\lambda)$ and (2.8) imply
that $\langle w_0 \rangle = (de(\lambda)/ d\lambda)_{|\lambda=0} \geq
0$.  This is analogous to the result \cite{DR} that the mean phase space
volume contraction in the stationary state is nonnegative.

To prove (2.8) we first define
\begin{equation}
g(\sigma,t) = {\Bbb{E}}_{\sigma}[e^{-\lambda W(t)}]
\label{2.15}
\end{equation}
as the expectation value of
$e^{-\lambda W(t)}$ conditioned on the system being in state $\sigma$
at time $t = 0$.  We then have
\begin{eqnarray}
{d \over dt} g(\sigma,t) & =
& \sum_{\sigma^\prime} k(\sigma,\sigma^\prime) e^{-\lambda
w(\sigma,\sigma^\prime)} g(\sigma^\prime,t) - r(\sigma) g(\sigma,t)\cr
~~~~ & = & \sum_{\sigma^\prime}
k(\sigma, \sigma^\prime)^{1-\lambda} k(\sigma^\prime,\sigma)^\lambda
g(\sigma^\prime,t) - r(\sigma) g(\sigma,t)\cr
~~~~ & = & L_\lambda g(\sigma,t)\, .
\label{2.16}
\end{eqnarray}
Eq. (2.14) is to be solved subject to the initial condition $g(\sigma,0) = 1$.
Therefore
\begin{equation}
\langle e^{-\lambda W(t)} \rangle =
\sum_{\sigma} \mu_{s}(\sigma) g(\sigma,t) =
\sum_{\sigma,\sigma^\prime} \mu_{s}(\sigma)
(e^{L_\lambda t})_{\sigma \sigma^\prime}\, .
\label{2.17}
\end{equation}

Clearly $(e^{L_{\lambda}t)})_{\sigma \sigma'} > 0$ by
our assumption on $k(\sigma,\sigma')$
above.
The Perron-Frobenius theorem states that $L_\lambda$ has a unique maximal
eigenvector $f_{\lambda}$ which 
is characterized by $f_{\lambda}(\sigma) > 0$
and satisfies $L_{\lambda}f_{\lambda}(\sigma) =
- e(\lambda) f_{\lambda}(\sigma)$ with
real maximal $- e(\lambda)$. This implies the existence of the limit in (2.7).
Furthermore $L_{\lambda}$ and $L_{\lambda}^{\ast}$ have the same maximal
eigenvalue. Using now the definition (2.16) we see that
\begin{equation}
        L_{\lambda}^{\ast} = L_{1-\lambda}\, .
\label{2.18}
\end{equation}
Hence the maximal eigenvector $\bar{f}_{\lambda}$
of $L_{\lambda}^{\ast}$ satisfies
\begin{equation}
 L_{\lambda}^{\ast} \bar{f}_{\lambda}(\sigma) = -e(\lambda)
\bar{f}_{\lambda}(\sigma) = L_{1 - \lambda}
        \bar{f}_{\lambda}(\sigma)\, .
\label{2.19}
\end{equation}
Since $L_{1-\lambda}f_{1-\lambda}(\sigma) = - e(1 -
\lambda)f_{1-\lambda}(\sigma)$ and since $\bar{f}_{\lambda} > 0$,
we conclude by uniqueness that $\bar{f}_{\lambda} = f_{1-\lambda}$
and $e(\lambda) = e(1 - \lambda)$. \bigskip\\
{\large {\bf 2.2 Time Reversal}} \medskip\\
For a Markov chain with transition probability
$p(\sigma,\sigma^\prime)$ from $\sigma$ to $\sigma^\prime$
the history  $\{{\bf \sigma}\} = \{\sigma_{0},\ldots,\sigma_{n}\}$ 
has the probability
\begin{equation}
P(\{{\bf \sigma}\}) = \mu_s(\sigma_0) p(\sigma_0, \sigma_1)
... p(\sigma_{n-1},\sigma_n)
\label{2.20}
\end{equation}
when starting in the stationary measure $\mu_s$.  In analogy to (2.6) we
define  the
action functional as 
\begin{equation}
W(n,\{\sigma\}) =  -\sum_{j=1}^n
\log [p(\sigma_j,\sigma_{j-1})/p(\sigma_{j-1},\sigma_j)]\, .
\label{2.21}
\end{equation}
If we now denote the time reversed history of $\{\sigma\}$ by $R\{\sigma\}
= \{\sigma_n,...,\sigma_0\}$, then (2.19) can be rewritten as 
\begin{eqnarray}
W(n,\{\sigma\}) & = &
- \log[P(R\{{\bf
\sigma}\})/P(\{\sigma\})] +  \log[\mu_s(\sigma_n)/\mu_s(\sigma_0)] \nonumber
\\  & 
= & \bar W(n,\{\sigma\}) +  \log[\mu_s(\sigma_n)/\mu_s(\sigma_0)] \, .
\label{2.22}
\end{eqnarray}
One now immediately verifies that $\langle e^{-\lambda \bar W(n)} \rangle =
\langle e^{-(1-\lambda)\bar W(n)} \rangle$, where the average is with
respect to $P(\{{\bf \sigma}\})$.  Thus for $\bar W(n)$ the fluctuation
theorem  holds even for finite $n$.

The form (2.20) is analogous to the form which the GC functional takes for
thermostatted systems when the latter deterministic dynamical evolution is
described via Markov partitions and it shows explicitly the role of time
reversal in the original GC theorem.  This observation forms the basis of
the analysis of C. Maes in \cite{M}, where he studies the fluctuation
theorem for space-time Gibbs measures and also discusses symmetry
transformations different from time reversal.

For the time continuous jump process a path $\{ \sigma_s, 0 \leq s \leq
t\}$ is time reversed as $\{ \sigma_{t-s}, 0 \leq s \leq t\}$.  The time
reversed process has the jump rates $k^R(\sigma,\sigma^\prime) =
k(\sigma^\prime,\sigma) \mu_s(\sigma^\prime)/\mu_s(\sigma)$ whereas the
inverse waiting times, $r(\sigma)$, and the steady state, $\mu_s(\sigma)$,
remain unmodified.  If $P_{[0,t]}$ denotes the path measure of the
stationary process in the time window $[0,t]$ and $P^R_{[0,t]}$ the one of the
corresponding time reversed process, then $P^R_{[0,t]}$ has a density
relative to $P_{[0,t]}$ and
\begin{equation}
P^R_{[0,t]} = e^{-W(t)} (\mu_s(\sigma_t)/\mu_s(\sigma_0)) P_{[0,t]}
\label{2.23}
\end{equation}
with $W(t)$ from (2.4).  Thus, up to boundary terms, the action functional
equals $-\log(d P^R_{[0,t]}/d P_{[0,t]})$ with $dP^R/dP$ denoting the
Radon-Nikodym deri\-vative.  

(2.21) remains meaningful for a stochastic dynamics which is {\it not
Markov} and it can thus be used as a definition of the action functional in
such a more general context.

One may wonder whether time-reversal could be replaced by some other
transformation on path space.  An obvious candidate is an internal symmetry
$S : {\cal S} \to {\cal S}$ such that $S \circ S = 1$.  Using
this transformation in (2.21) the GC action functional becomes
\begin{equation}
W(t) = -\int_0^t ds \log[r(S \sigma_s)/r(\sigma_s)]
\label{2.24}
\end{equation}
up to boundary terms.  From the examples we have studied it seems that only
time-reversal leads to an action functional which has a simple physical
interpretation.  \bigskip\\
{\large {\bf 2.3 Sum of several generators}}\medskip\\
We can generalize the above analysis to the case where 
the generator in (2.1) is the sum of several generators:  
$L = \sum_{j=1}^{m}L^{(j)}$ with
\begin{equation}
L^{(j)} f(\sigma) = \sum_{\sigma'} k^{(j)}(\sigma,\sigma') 
[f(\sigma') - f(\sigma)]\, .
\label{2.25}
\end{equation}
We require that for transitions between a given pair $\sigma,\sigma'$ at
most one of the rates $k^{(j)}$ is different from zero.  It follows then
from our assumptions on $k(\sigma,\sigma^\prime)$ that if
$k^{(j)}(\sigma,\sigma') > 0 $, then also $ k^{(j)}(\sigma',\sigma) >
0$. With this decomposition we define, in complete analogy with
(2.4)--(2.6), the $j$-th action functional
\begin{equation}
        W^{(j)}(t,\{\sigma_s, 0\leq s\leq t\}) = \int_{0}^t
        \sum_{\sigma,\sigma'} w^{(j)}_{\sigma,\sigma'}(s) ds
\label{2.26}
\end{equation}
and the logarithm of the generating function
\begin{equation}
e(\lambda_{1},\ldots,\lambda_{m}) =  
\lim_{t \rightarrow \infty} - {1 \over t} \log \langle \exp[-
\sum_{j=1}^{m}\lambda_{j} W^{(j)}(t)] \rangle \, .
\label{2.27}
\end{equation}
Repeating the arguments of Section 2.1 we conclude that 
the fluctuation theorem now takes the form 
\begin{equation}
e(\lambda_{1},\ldots,\lambda_{m}) =  
e(1 - \lambda_{1},\ldots,1 - \lambda_{m})\, .
\label{2.28}
\end{equation}
The corresponding analog of (2.10) is
\begin{equation}
\hat e(w_{1}, \ldots, w_{n}) - \hat e(-w_{1}, \ldots,
-w_{m}) = -\sum_{j=1}^m w_{j}
\label{2.29}
\end{equation}
We shall use the general form (2.26) in Section 8 to prove the Onsager
relations but note here that there are many ways of splitting $L$ into a
sum of $L^{(j)}$'s.  In fact we can choose a different $L^{(j)}$ for every
pair of points $(\sigma, \sigma^\prime)$ for which $k(\sigma,\sigma^\prime)
\ne 0$.  In all cases setting $\lambda_{j} = \lambda$ for all $j$ we
recover (2.8) and (2.10) for $W = \sum_{j=1}^{m} W^{(j)}$.  We do not
expect in general that the distribution of $(1/t) W^{(j)}(t)$ separately
satisfies any fluctuation theorem.  On the other hand we can consider, as a
degenerate case of (2.23), a situation in which our system is composed of
several independent systems with generators $L^{(j)}$.  In that case, of
course, each $L^{(j)}$ satisfies the symmetry relation, 
$e_{j}(\lambda_{j}) = e_{j}(1-\lambda_{j})$, $\hat e_{j}(w_{j}) -
\hat e_{j}(-w_{j}) = -w_{j}$,
and 
$e(\lambda_{1},\ldots \lambda_m) = \sum_{j=1}^m e_{j} (\lambda_{j})$, $\hat
e(w_{1},\ldots,w_{m}) = \sum_{j=1}^m \hat e_{j} (w_{j})$.
\bigskip\\ 
{\large {\bf 2.4 Entropy production}}\medskip\\ 
For the probability distribution
$\mu(\sigma,t)$ at time $t$ the Gibbs entropy is, as usual, given by 
\begin{equation}
S_G(\mu(t)) = -\sum_{\sigma} \mu(\sigma,t) \log \mu(\sigma,t)\, .
\label{2.30}
\end{equation}
Using (2.3) the rate of change of $S_G$ can be written in the form
\begin{equation}
{d \over dt} S_G(\mu(t)) = R(\mu(t)) - A(\mu(t))\, ,
\label{2.31}
\end{equation}
where
\begin{equation}
R(\mu(t)) = {1 \over 2} \sum_{\sigma,\sigma^\prime}
[k(\sigma,\sigma^\prime)\mu(\sigma,t) -
k(\sigma^\prime,\sigma)\mu(\sigma^\prime,t)]\log[{\mu(\sigma,t)
k(\sigma,\sigma^\prime)
\over \mu(\sigma^\prime,t) k(\sigma^\prime,\sigma)}] \geq 0
\label{2.32}
\end{equation}
and
\begin{equation}
 A(\mu(t)) = \langle I\rangle_{\mu(t)}, \quad
I(\sigma) = \sum_{\sigma^\prime} k(\sigma,\sigma^\prime) \log [{k(\sigma,
\sigma^\prime) \over k(\sigma^\prime, \sigma)}] \, .
\label{2.33}
\end{equation}
(When the generator is a sum, $L = \sum_{j=1}^{m}L^{(j)}$, as 
in the previous section, then the rate of change in Gibbs entropy is also
a sum over $j$ and consequently $R = \sum_{j=1}^{m}R^{(j)}$,
$A = \sum_{j=1}^{m}A^{(j)}$).

The breakup in (2.30) has the property that $R$ is non-negative, while $A$
is linear in $\mu$.  This suggests that we identify $R$ as the entropy
produced by the stochastic jumps and $A$ with the entropy flow which can
have either sign.  Defining now the ``integrand'' in $A$, $I(\sigma)$, as a
``microscopic'' entropy flux we note that the expectation value of $W(t)$,
starting with some initial distribution $\mu(\sigma,0)$, is given by
\begin{equation}
\langle W(t)\rangle_{\mu(0)} = \int_{0}^{t} ds  \langle I \rangle_{\mu(s)}\, .
\label{2.34}
\end{equation}
We can therefore say that the average action equals
the entropy flow integrated over the time span $t$.
There is of course some arbitrariness
in this identification.  We could for example add some positive linear term
to $R$ and subtract it from $A$.  Our choice of  
splitting is however quite natural in the examples discussed below.

When the system is in the steady state $\mu_{s}$, 
${d \over dt}S_{G} (\mu_s) = 0$ and the entropy flow balances the entropy
production, i.e.
\begin{equation}
R(\mu_{s}) = A(\mu_{s}).
\label{2.35}
\end{equation}
Thus
\begin{equation}
\langle t^{-1} W(t) \rangle = R(\mu_{s}) = A(\mu_{s})
\label{2.36}
\end{equation}
and hence the average action equals the rate of entropy production in the
steady state.  This suggests to identify $W$ with the microscopic
production of Gibbs entropy, a role played by the phase space contraction
in the context of thermostatted systems; see also Section 7.

We remark that $R(\mu_s) = A(\mu_s) = 0$ in case $k(\sigma,\sigma')$
satisfies detailed balance with respect to the invariant measure
$\mu_{s}(\sigma) \sim e^{-V(\sigma)}$, i.e.
\begin{equation}
        e^{-V(\sigma)} k(\sigma,\sigma') = e^{-V(\sigma')} 
        k(\sigma',\sigma)\, .
\label{2.37}
\end{equation}
Then each term in the sum defining $R(\mu_s)$ in (2.30) vanishes.  This
implies by (2.29) that $A(\mu_s) = \langle I \rangle _{\mu_{s}} = 0$. The
action functional (2.6) now becomes a telescoping sum with
\begin{equation}
        W(t) = V(\sigma_{0}) - V(\sigma_{t})\, .
\label{2.38}
\end{equation}
It then follows from
(2.7) that $e(\lambda) = 0$
and the fluctuation theorem becomes empty.
 More generally if
\begin{equation}
        k(\sigma,\sigma') = k_{0}(\sigma,\sigma') k_{1}(\sigma,\sigma')
        \label{2.39}
\end{equation}
and $k_{0}(\sigma,\sigma')$ satisfies detailed balance while $k_{1}$ is a
nonequilibrium driving ``force'', then the $\log k_{0}$ terms of $W(t)$ sum
to a boundary value as before and only the weight corresponding to $k_{1}$
contributes in $W(t)$. We will encounter such a situation for driven
lattice gases.

\section{Bulk driven lattice gases}
\setcounter{equation}{0}

An important example of an SNS with spatial structure is a stochastic
lattice gas where particles jump at random to neighboring lattice sites. We
envisage two mechanisms for how the conserved particle density is driven
away from equilibrium. The first one, studied in this section, is a global
driving force which corresponds to driven diffusive systems (DDS). In the
second one, considered in the next section, particles are injected/removed
at the boundaries whereas the bulk is governed by reversible dynamics
satisfying detailed balance with respect to an equilibrium stationary
measure.  In both cases the full dynamics does not satisfy detailed balance
with respect to the stationary measure $\mu_{s}$.

We consider particles with exclusion hopping on a regular lattice
$\Bbb{Z}$$^{d}$.  The configuration is specified by the occupation
variables $\eta_{x}, x \in \Bbb{Z}$$^{d}$, taking the values $0$ (empty)
and $1$ (occupied). Particles on different sites may interact by making the
jump rate of a given particle depend on the configurations on nearby
sites. As explained in Section 2, the fluctuation theorem holds in
general. For notational simplicity we restrict ourselves however to the
one-dimensional lattice, $d = 1$.

We first study particles subject to a uniform drive. This will set up a
steady state current in a finite system of particles jumping on a ring, $x
= 1,\ldots,\ell$, with $x = \ell +1$ identified with $x = 1$. A particle
configuration is denoted by $\eta = (\eta_{1},\ldots, \eta_{\ell})$.  We
denote by $c_{xx+1}(\eta)$ the exchange rate for the pair of sites $(x,
x+1)$, i.e., if $\eta_{x} = 1$ $(0)$ and $\eta_{x+1} = 0$ $(1)$, then
$c_{xx+1}(\eta)$ is the rate at which the particle at $x$ $(x+1)$ jumps to
the empty site at $x+1$ $(x)$.

In equilibrium the distribution of particles is determined by the Gibbs
measure $\mu_{eq} \sim \exp[-\beta H]$ where $\beta$ is the inverse
temperature and $H(\eta)$ is the energy function.  The rates $c_{xx+1}$
satisfy the condition of detailed balance, with respect to $\mu_{eq}$
whenever
\begin{equation}
        e^{-\beta H (\eta)} c_{xx+1}(\eta) = e^{-\beta H(\eta
        ^{xx+1})} c_{xx+1}
        (\eta^{xx+1})\, .
\label{3.1}
\end{equation}
Here $\eta^{xx+1}$ denotes the configuration $\eta$ with the occupations at
$x,x+1$ interchanged.  A simple example of $H(\eta)$ is the energy of a
system with nearest neighbor interactions
\begin{equation}
H(\eta) = -J \sum_{x=1}^{\ell}\eta_{x}\eta_{x+1}\, ,
\label{3.2}
\end{equation}
but such an explicit form is not used here.  Rates satisfying (3.1) lead
the system to equilibrium.  In order to drive the lattice gas by a uniform
external force field of strength $F$ we add to $H(\eta)$ the linear term $-
\sum_{x} F x \eta_{x}$. While this term does not respect the boundary
conditions, energy differences always do. Therefore the rates for a uniform
drive are assumed to satisfy
\begin{equation}
        c^{F}_{xx+1}(\eta) = c^{F}_{xx+1}(\eta^{xx+1}) \exp[-
\beta(H(\eta^{xx+1} 
        )-H(\eta) - F(\eta_{x} - \eta_{x+1}))]\, .
        \label{3.3}
\end{equation}
We have called this condition local detailed balance \cite{KLS,ELS1},
because locally a particle feels a linear potential. It is only through the
periodic boundary condition, that the steady state becomes a nonequilibrium
one.  For a closed system the particles would pile up at the right (for $F
> 0$).

To state the fluctuation
theorem we merely have to apply the results of Section 2.
The logarithmic ratio in (2.5) is
\begin{equation}
        - \beta(H(\eta^{xx+1}) - H (\eta)) + \beta F(\eta_{x} - \eta_{x+1})
        \label{3.4}
\end{equation}
with $\eta$ the configuration before and $\eta^{xx+1}$ the configuration
after the jump. The energy difference summed over the jumps is telescoping
and yields a pure surface term as in (2.36). Therefore the relevant part of
the action functional is given by

\begin{equation}
        W(t) = \beta F \int^t_{0} \sum_{x=1}^{\ell} J_{x,x+1} (s) 
        ds\, . 
        \label{3.5}
\end{equation}
$J_{x,x+1}(s)$ is the actual particle current across the bond $(x,x+1)$,
i.e.  for a given history $\{\eta(s), 0 \leq s \leq t\}$, $J_{x,x+1}(s)$ is
a sequence of $\delta$-functions located at the times of jump through the
bond $(x,x+1)$ with weight $+1$ $(-1)$ if the jump is from $x$ $(x+1)$ to
$x+1$ $(x)$.  The action functional is $\beta F$ times the the total
current integrated over the time span $t$, which equals the signed number
of all jumps up to time $t$.

The entropy production $R(\mu)$ can be read off from (2.30).
If we set $\mu(\eta) = g (\eta) Z^{-1} e^{-\beta H(\eta)}$
and denote by $\langle \cdot \rangle_{eq}$
the average over $Z^{-1} e^{- \beta H(\eta)}$, then
\begin{eqnarray}
R(\mu) & = &{1 \over 2} \sum^{\ell}_{x = 1}
\langle c_{xx+1}^{F} (\eta) [e^{-\beta
F(\eta_{x} - \eta_{x+1})} g(\eta^{xx+1}) - g(\eta)] \nonumber \\
& & [\log e^{-\beta F(\eta_{x}-\eta_{x+1})} g(\eta^{xx+1}) - 
\log g(\eta)]\rangle_{eq}\, .
        \label{3.8}
\end{eqnarray}
We clearly have $R(\mu) > 0$ unless $F = 0$ \cite{S}.
The  identification of $R(\mu)$ as entropy production
is further supported by taking
$\mu$ to be a state of local equilibrium. In the limit of slow density
variations $R$  coincides with the phenomenological entropy
production based on the nonlinear diffusion equation as a macroscopic
equation for the density \cite{S}.

We emphasize that the action functional is equal to the total current only
if the condition (3.3) of local detailed balance holds. To illustrate this
point we consider nearest and next nearest neighbor jumps. Local detailed
balance for next nearest neighbor jumps means
\begin{equation}
        c^{F}_{xx+2}(\eta) = c^{F}_{xx+2}(\eta^{xx+2}) \exp[-
\beta(H(\eta^{xx+2} 
        )-H(\eta) - 2F(\eta_{x} - \eta_{x+2}))]\, .
        \label{3.9}
\end{equation}
As before, $W(t)$ constructed according to the rule (2.5) equals the total
current. There are many ways to violate (3.7). Just as an example, if in
(3.7) we replace $2F$ by $\alpha F$, then the action functional is $W(t) =
\beta F \int dt \sum_{x}(J_{x,x+1}(t)$ $ + \alpha J_{x,x+2}(t))$ which equals
the particle current only if $\alpha = 2$. Thus, while the fluctuation
theorem holds in great generality, the specific physical interpretation is
linked to local detailed balance.  We could also in this case consider the
generator as the sum of two generators, one for nearest and one for 
next nearest neighbor jumps, and have more general fluctuation
theorems as in Section 2.3.  The physical interpretation would however remain
obscure.

\section{Boundary driven lattice gases}
\setcounter{equation}{0}

We use the same set-up as in the previous section. Only now the box
consisting of sites $\{1,2,\ldots,\ell\}$ is not periodic, and there is no
driving force. Instead particles are injected and removed at the left $(x =
1)$ and right $(x = \ell)$ boundary. The generator, $L$, for the full
dynamics is naturally decomposed as a sum of three pieces,
\begin{equation}
        L = L_{b} + L_{1} + L_{\ell}\, .
        \label{4.1}
\end{equation}
 $L_{b}$ is the dynamics inside the box satisfying detailed balance with
respect to $e^{- \beta H}$, cf. (3.1) with $x=1,\ldots,\ell-1$. $L_{b}$
conserves the number of particles.  $L_{i}$ models the particle source/sink
at the right and left boundaries and has the form
\begin{equation}
        L_{i} f(\eta) = c_{i}(\eta) [f(\eta^i) - f(\eta)]\, ,\quad i = 1, 
		\ell \, .
        \label{4.2}
\end{equation}
$c_{i}(\eta)$ is the rate for the
transition $\eta_{i}$ to $1-\eta_{i}$
while the remaining configuration is untouched. The configuration after this
transition is denoted by $\eta^i$. 
As  before we will require the rate $c_i(\eta)$ to satisfy 
local detailed balance in the form \cite{ELS1} 
\begin{equation}
        c_{i}(\eta) = c_{i}(\eta^i) \exp [- \beta (H(\eta^i) - H(\eta)) + \beta
        \gamma_{i} (1-2\eta_{i})]\, , 
        \label{4.3}
\end{equation}
where local refers now to the boundary points which are coupled to
reservoirs with chemical potentials $\gamma_{1}$ and $\gamma_{\ell}$. Thus
if we consider the dynamics generated only by $L_{b}+L_{i}$, then the
dynamics will satisfy detailed balance with respect to the stationary
distribution $Z^{-1} \exp[- \beta H(\eta) +\beta \gamma_{i} N(\eta)]$
(achieved in the long time limit) with $N(\eta) = \sum^\ell_{x=1} \eta_{x}$
the number of particles.  The same will be true for the full $L$ in (4.1),
if $\gamma_1 = \gamma_{\ell}$.  If on the other hand $\gamma_{1} \neq
\gamma_{\ell}$, then the sources are unbalanced and there will be in the
steady state a net flux of particles through the system going from the
reservoir with high to the reservoir with low chemical potential, i.e.\
from left to right if $\gamma_1 > \gamma_{\ell}$.

We now apply the results of Section 2 (with $\lambda_j = \lambda$). The
logarithmic ratio (2.5) reads
\begin{equation}
- \beta (H(\eta^{xx+1}) - H(\eta)) - \beta \sum_{i=1, \ell} (H(\eta^i) -
H(\eta)) + 
         \sum_{i=1, \ell} \beta \gamma_{i} (N(\eta^i) - N(\eta)),
        \label{4.4}
\end{equation}
where $\eta$ is the configuration before and $\eta^{xx+1}$, resp. $\eta^i$,
the configuration after the jump. Again the energy differences sum to a
surface term. Let $J_{i}(t)$ be the boundary current, $i=1, \ell$. It is a
sequence of delta functions located at times when $\eta_{i}$ jumps to
$1-\eta_{i}$ with weight $1$ $(-1)$ for the transition from $0$ to $1$
(from $1$ to $0$). When the boundary current $J_1$ is positive there is a
net flux of particles from the $i$th reservoir into the system. Then
\begin{equation}
        W(t) = \beta \int^t_{0} (\gamma_{1} J_{1}(s) +
		\gamma_{\ell} J_{\ell}(s))ds\, .
        \label{4.5}
\end{equation}

Because of particle conservation
there are many equivalent expressions for $W(t)$ in the sense that they
differ only by surface terms and have therefore the same rate function.
Let us denote by $N_{[x,y]} (t), \ x \le y$, the number of particles in
the interval $[x,y]$ at time $t$. By the conservation law for the number of
particles we have
\begin{equation}
N_{[x,y]}(t) - N_{[x,y]}(0) = \int^t_{0}(J_{x-1,x}(s) - J_{y,y+1}(s)) 
ds\, .
        \label{4.6}
\end{equation}
Since $N_{[x,y]}(t) \le |y-x|+1, \int^t_{0} J_{x-1,x} (s)ds$ and
$\int^t_{0} J _{y,y+1} (s)ds$ must have the same large deviations. If we
include in the mass balance also the boundary currents, then we can replace
in (4.5) $J_{1}(t)$ by any $J_{x,x+1}(t)$ and $J_{\ell}(t)$ by any
$-J_{x,x+1}(t)$ without modifying the rate function $\hat{e}(w)$. While
this gives many equivalent choices, the simplest one is perhaps
\begin{equation}
W_{x}(t) =      \beta(\gamma_{1}-\gamma_{\ell}) \int^t_{0} J_{x,x+1}(s)ds =
\beta(\gamma_{1}-\gamma_{\ell}) \tilde{W}_{x}(t)\, .
        \label{4.7}
\end{equation}
 $W_{x}(t)$ differs from $W(t)$ only by a surface term. The action
functional $\tilde{W}_{x}(t)$ is the time integrated particle current
across the bond $(x,x+1)$.  The generating function for
$\tilde{W}_{x}(t)$ satisfies the fluctuation theorem in the form
$\tilde{e}(\lambda) = \tilde{e}(\beta(\gamma_{1}-\gamma_{\ell}) -\lambda)$,
with $\tilde{e}$ independent of the choice of $x$.

The conservation law (4.6) can be used also in the driven lattice gas
of the previous section to produce
equivalent action functionals. For example $W(t)$ of (3.5) is equivalent to
\begin{equation}
\bar{W}(t) = \ell \beta F \int_{0}^{t} J_{x,x+1}(s) ds \, .  \label{4.8}
\end{equation}
However the symmetric form (3.5) seems to be more accessible to analytic
computations, as can be seen from the examples described in the Appendix.

\section{Fluctuation theorem for diffusion processes}
\setcounter{equation}{0}

We consider next a general diffusion process $x_{t} \in \Bbb{R}$$^{n}$. It
is specified by a drift vector $c(x)$ (a vector field on $\Bbb{R}$$^{n}$)
and a positive definite diffusion matrix $a(x) > 0$ (an $n \times n$ matrix
valued function on $\Bbb{R}$$^{n}$).  For later convenience we write the
generator as
\begin{equation}
        L = {1 \over 2} \nabla \cdot (a \nabla) + c \cdot \nabla
        \label{6.1}
\end{equation}
with $\nabla$ representing differentiation with respect to $x$. $x_{t}$ is
the solution of the stochastic differential equation
\begin{equation}
        d x_{t} = (c + {1 \over 2} (\nabla \cdot a))(x_{t})dt +
        \sqrt{a}(x_{t}) db(t)
        \label{6.2}
\end{equation}
with $db(t)/dt$ standard white noise. We use Ito's convention for
stochastic differentials.

We claim that for diffusion processes the action functional $W(t)$ is given
by
\begin{equation}
 W(t) = 2 \int_{0}^t (a^{-1} c(x_{s})) \cdot dx_{s} + \int^t_{0} a \nabla
(a^{-1} 
 c) (x_{s}) ds\, .
        \label{6.3}
\end{equation}
We may think of $W(t)$ as the limit of the symmetrized
(Stratonovich) approximation,
\begin{equation}
        W(t) = \lim_{\epsilon \rightarrow 0} \epsilon 
		\sum_{j=0}^{t/\epsilon} (a^{-1}
        c(x_{(j+1)\epsilon}) + a^{-1} c (x_{j \epsilon})) 
		\cdot (x_{(j+1) \epsilon} -
        x_{j \epsilon})\, .
        \label{6.4}
\end{equation}
If $a$ is diagonal and independent of $x$, then $a_{ii}/2$ corresponds
physically to the temperature $\beta^{-1}$.  In this case $W(t)/ \beta$
equals the work done by the force $c$ on the system during the time span
$t$, in accordance with the physical meaning of the action functional for
lattice gases.

As before, we expect that the equilibrium forces do not show up in the
large deviations of $W(t)$. For diffusion processes detailed balance with
respect to $e^{-U(x)}$ means that
\begin{equation}
        c = - {1 \over 2} a \nabla U\, .
        \label{6.5}
\end{equation}
Inserting into (5.3), by Ito's lemma \cite{SV},
\begin{equation}
        - \int^{t}_{0} \nabla U(x_{s}) \cdot dx_{s} - {1 \over 2} 
		\int^t_{0} a \nabla
        \nabla U(x_{s}) ds = U(x_{0}) - U(x_{t})\, ,
        \label{6.6}
\end{equation}
which is indeed a pure surface term. Only driving forces make a 
contribution to $W(t)$
proportional to $t$.

Since $a > 0$ by assumption, we are in the uniformly elliptic case, where
the transition probability, $e^{Lt}$, has a density and $(e^{Lt})_{x,x'} >
0$ for $t > 0$. We require that the drift $c$ is sufficiently
confining. Then $x_{t}$ has a unique stationary measure with density
$\mu_{s}(x) > 0$, satisfying $\mu_s e^{Lt} = \mu_s $.

Let us denote by $\langle \cdot \rangle$ expectation with respect to the
stationary process $x_{t}$ (starting in the invariant measure
$\mu_{s}$). Then
\begin{equation}
        \lim_{t \rightarrow \infty} - {1 \over t} \log \langle 
		e^{-\lambda W(t)} \rangle
        = e(\lambda)
        \label{6.7}
\end{equation}
and the fluctuation theorem
\begin{equation}
        e(\lambda) = e(1-\lambda)
        \label{6.8}
\end{equation}
should hold.

To prove (5.8), as for jump processes, we first define the function
$g(x,t) = {\Bbb{E}}_{x}(\exp [- \lambda W(t)])$
as the expectation with respect to the stationary
diffusion process $x_{t}$ conditioned to start at $x$.
Then $g(x,t)$ satisfies
\begin{equation}
{\partial \over \partial t} g(x,t) = L_{\lambda}g(x,t)
\label{6.9}
\end{equation}
with
\begin{equation}
        L_{\lambda} = {1 \over 2} \nabla \cdot (a \cdot \nabla) + c \cdot
\nabla 
        - 2 \lambda c \cdot \nabla - \lambda (\nabla \cdot c) -
        2 \lambda(1- \lambda) c \cdot a^{-1} c\, ,
\label{6.10}
\end{equation}
where the last two terms act as multiplication operators. We conclude that
\begin{equation}
        \langle e^{-\lambda W(t)}\rangle = \int \int \mu_{s}(x) 
		(e^{L_{\lambda}t})_{x,x'} dx dx'.
\label{6.11}
\end{equation}
On the other hand, from (5.10),
\begin{equation}
        L^\ast_{\lambda} = L_{1- \lambda}\, ,
\end{equation}
which implies (5.8) by Perron-Frobenius.

Eqs. (5.10) and (5.11) can also be viewed as an application of the
Cameron-Martin-Girsanov formula \cite{SV}. The path measure of the process
generated by $L_{\lambda}$ has a density relative to the path measure of
the process generated by $L$. We write this density as $\exp[R]$.  Then,
according to \cite{SV}, Section 6.4,
\begin{eqnarray}
R & = & -2 \lambda \int a^{-1} c(x_{t}) \cdot dx_{t} + 2 \lambda \int (a^{-1}c)
\cdot (c + {1 \over 2} \nabla a) (x_{t}) dt \nonumber \\
& & - 2 \lambda^2 \int c \cdot a^{-1} c(x_{t}) dt - 2 \lambda (1-\lambda) \int
c \cdot a^{-1} c (x_{t}) dt - \lambda \int \nabla \cdot c(x_{t}) dt
\nonumber \\ 
& = & - 2 \lambda \int a^{-1}
c(x_{t}) \cdot dx_{t} - \lambda \int a \nabla (a^{-1}c)(x_{t}) 
dt\nonumber\\
& = & -\lambda W(t)\, , 
\label{6.12}
\end{eqnarray}
where we used the identity
\begin{equation}
        a(\alpha) {d \over d \alpha} a^{-1} (\alpha) = - \left( {d \over d
\alpha} 
        a(\alpha) \right) a^{-1} (\alpha)
        \label{6.13}
\end{equation}
valid for a parameter-dependent matrix $a(\alpha)$.

{}From the argument of Section 2, we conclude that $W(t)/t$ has large
deviations as stated in (2.11) with
a rate function satisfying the symmetry (2.10).

\section{Particle under mechanical and thermal drive}
\setcounter{equation}{0}

In the stochastic differential equations of Section 5 the diffusion acts
everywhere on the configuration space ${\Bbb{R}}^{n}$. This is appropriate
for strongly overdamped stochastic systems such as those considered in
time-dependent Ginzburg-Landau theories.  In many physical situations,
however, stochasticity is assumed to act only in velocity space.  Then the
matrix $a$ in (5.1) has zero eigenvalues and, the results of Section 5 are
not directly applicable.  To understand the required modifications we
consider, as an example, a mechanical particle of mass $m$ with position
$x$ in the $d$-dimensional torus $T^d$ and velocity $v$ in $\Bbb{R}$$^{d}$
subject to noise and friction. The particle has the mechanical energy $H =
{1 \over 2} mv^2 + U(x)$, where $U$ is some periodic potential. In
equilibrium the stationary distribution is given by $\exp [- \beta H(x,v)]$
with $\beta^{-1}$ the temperature.  We envision two mechanisms to drive the
system out of equilibrium: (i) There is a mechanical driving force
$F(x)$. The standard example is a constant electric field $E$, i.e. $F(x) =
E$.  If $F$ has a periodic potential part, it may be added to $U$ but as we
will see this has no effect on the large deviation of the action
functional.  (ii) There is a nonuniform temperature $\beta(x)^{-1}$. The
case of a periodic variation in $\beta$ (and constant friction) has been
studied recently in the context of Brownian motors \cite{B,BB} with the at
first sight surprising result that the steady state maintains a nonzero
current.
The dynamics of the particle is
governed by the stochastic differential equation
\begin{equation}
m{d^2 \over dt^2} x_{t} = - \nabla U(x_{t}) + F(x_{t}) -
m \gamma (x_{t}) v _{t} + (2m
        \gamma(x_{t})/\beta(x_{t}))^{1 \over 2} \xi(t)\, .
        \label{7.1}
\end{equation}
Here $\gamma (x) > 0$ is the friction coefficient, $\beta(x) > 0$ the inverse
temperature and $\xi(t)$ standard white noise. The generator of
the corresponding Fokker-Planck equation reads
\begin{equation}
L = v \cdot \nabla_{x} + \left( - {1 \over m} \nabla U +{1 \over m} F - \gamma
(x) v \right) \cdot \nabla_{v} + (\gamma(x)/m \beta(x)) 
\nabla^2_{v}\, .
        \label{7.2}
\end{equation}
We claim that the action functional for the fluctuation theorem is given by
\begin{equation}
        W(t) = \int_{0}^t \beta(x_{s}) F(x_{s}) \cdot dx_{s} + \int_{0}^t
H(x_{s}, 
        v_{s}) d \beta(x_{s})
        \label{7.3}
\end{equation}
for $x_{s}, v_{s}$ a solution to (6.1). The first term is $\beta
\times$(work done by $F$) and the second term represents $\beta
\times$(work due to thermal gradients) integrated along the trajectory
$x_{s}$, $0 \leq s \leq t$.  If $F = -\nabla V$, then up to a surface term
\endgraf \noindent $\int_{0}^{t} \beta(x_{s}) F(x_{s}) \cdot dx_{s} =
\int_{0}^{t} V(x_{s}) d \beta(x_{s})$ which, in (6.3), should be added to
the energy $H(x,v)$.

Let $\langle \cdot \rangle$ denote the stationary average for (6.1). Then
\begin{equation}
        \lim_{t \rightarrow \infty} - {1 \over t} \log \langle 
		e^{- \lambda W(t)} \rangle
        = e(\lambda)
        \label{7.4}
\end{equation}
and $-e(\lambda)$ is the maximal eigenvalue of
\begin{equation}
        L_{\lambda} = L - \lambda(\beta(x) F(x) \cdot v + H(x,v) v 
		\cdot \nabla_{x}
        \beta(x))\, ,
        \label{7.5}
\end{equation}
the last two terms being considered as multiplication operators.
The corresponding
maximal eigenvector is denoted by $f_{\lambda}(x,v) > 0$. Let $R$ be the
velocity reversal operator, $R f(x,v) = f(x,-v)$. Then
\begin{equation}
R e^{-\beta(x)H} L_{\lambda} e^{\beta(x) H} R^{-1} R e^{- \beta(x)H}
f_{\lambda} 
(x,v) = -e(\lambda) R e^{-\beta (x) H} f_{\lambda}(x,v)\, .
        \label{7.6}
\end{equation}
By a straightforward computation
\begin{equation}
R e^{-\beta(x)H} L_{\lambda} e^{\beta(x) H} R^{-1} = 
L^\ast_{1-\lambda}\, .
        \label{7.7}
\end{equation}
Therefore
\begin{equation}
        L^\ast_{1-\lambda} R e^{-\beta H} f_{\lambda} = -e(\lambda) 
		R e^{-\beta H}
        f_{\lambda} = -e (1-\lambda) R e^{-\beta H} f_{\lambda}
        \label{7.8}
\end{equation}
by Perron-Frobenius, since $R e^{-\beta(x)H} f_{\lambda}(x,v) > 0$. We conclude
that the fluctuation theorem holds and $e(\lambda) = e(1-\lambda)$.

As for lattice gases, we expect that the action functional $W(t)$
is linked to the entropy balance. To spell out the details we consider the
time-dependent probability density $\rho_{t}(x,v)$ of
$(x_{t},v_{t})$. It satisfies
${\partial \over \partial t} \rho_{t} = L^\ast \rho_{t}$.
As usual the Gibbs entropy is given by
$S_{G}(\rho_{t}) = - \int \rho_{t} \log \rho_{t} dxdv$
and changes in time as
\begin{equation}
        {d \over dt} S_{G}(\rho_{t}) = \sigma(\rho_{t}) + 
        j_{s}(\rho_{t})\, .
        \label{7.9}
\end{equation}
The first term
\begin{equation}
        \sigma(\rho_{t}) = \int \gamma {1 \over \rho _{t}} 
		[\sqrt{m \beta} v \rho_{t}
        + {1 \over \sqrt{m \beta}} \nabla_{v} \rho_{t}]^2 dxdv
        \label{7.10}
\end{equation}
is positive definite. We identify it as the entropy production in the system.
The remainder reads
\begin{equation}
        j_{s}(\rho_{t}) = {d \over dt} \int \rho_{t} \beta H dxdv - 
		\int \rho_{t}
        \beta F \cdot v dxdv -
        \int \rho_{t} H v \cdot \nabla_{x} \beta dxdv\, .
        \label{7.11}
\end{equation}
We regard it as the entropy flow from the system to the mechanical and heat
reservoirs.

According to the rules used before, we construct $W(t)$ by integrating
the linear functional
defining the entropy flow along a stochastic trajectory. Using (6.11)
this yields
\begin{equation}
        \int^{t}_{0} \left\{ -{d \over ds} \beta H(x_{s},v_{s}) + \beta
        F(x_{s}) \cdot
        v_{s} + H(x_{s},v_{s}) v_{s} \cdot \nabla \beta (x_{s}) \right\} 
        ds\, .
\label{7.12}
\end{equation}
Since the first summand is a surface term, we recover (6.3).

\section{Stochastic and thermostatting heat reservoirs}
\setcounter{equation}{0}

Physical systems with stationary flows of heat, momentum, etc., are usually
modeled by coupling the system to thermal reservoirs represented by
stochastic forces acting near the boundaries \cite{BL}--\cite{GI}.  Away
from the boundary the time evolution obeys the same laws as an isolated
system.  To be specific, let us consider one such model of heat flow
through a classical system of $N$ particles of mass $m$ with position
$q_{j}$, momentum $p_{j}, j=1,\ldots,N$, inside the slab $\Lambda =
[-\ell -a, \ell +a] \times [0,\ell]^2$. Let the boxes $\Lambda_{-} = [-\ell - a,
-\ell] \times [0, \ell]^2$, $\Lambda_{+} = [\ell,\ell + a] \times [0,
\ell]^2$ be the left and right boundary zones.  The indicator functions of
these sets are denoted by $\chi_{\pm}$. The particles interact through a
short range pair potential $V$ and are confined to the slab by the wall
potential $V_{w}$. The equations of motion are
\begin{eqnarray}
{d \over dt} q_{j} & = & {1 \over m} p_{j}\, , \nonumber\\
{d \over dt} p_{j} & = & F_{j} + F_{w}(q_{j}) +
        \sum_{\delta = \pm} \chi_{\delta}
        (q_{j})(- \gamma p_{j} + (2m \gamma/\beta_{\delta})^{1/2} 
        \xi_{j}(t))\, ,
\label{8.1}
\end{eqnarray}
where $q_j \in \Lambda, p_j \in {\Bbb R}^3$, $F_{j}$ is the force acting on
the $j$-th particle, $F_{j} =$ \endgraf \noindent $- \sum^{N}_{i \neq j=1} \nabla
V(q_{j}-q_{i}), 
F_{w} = -\nabla V_{w}$ is the force from the wall. $\{\xi_{j}(t)\}$ are a
collection of independent white noises. The friction and the stochastic
forces operate only when the particle is in the boundary layers acting
there like thermal reservoirs at inverse temperature $\beta_{\delta}$.  If
$\beta_{+}= \beta = \beta_{-}$, then the steady state is $Z^{-1} \exp
[-\beta H], \ H = \sum^{N}_{j=1} ({1 \over 2m} p_{j}^2 + V_{w}(q_{j})) +
{1 \over 2}\sum^{N}_{i,j=1} V (q_{i}-q_{j})$. If $\beta_{+} \neq
\beta_{-}$, there is, in the steady state, a constant heat flux from the
higher to the lower temperature reservoir.  (This is a variation of the
model used to describe heat flow in a crystal where the Langevin 
forces act on the particles in the end layers \cite{LLR}, \cite{EPR}).

We now follow the method explained in Section 6. For the similarity
transformation we use
\begin{equation}
        \exp \left[ - \sum^{N}_{j=1} \beta (q_{j}) \left\{ {1 \over 2m} p_{j}^2
        + V_{w}(q_{j)} + {1 \over 2}
        \sum_{i=1}^{N} V (q_{i}-q_{j}) \right\} \right]\, .
        \label{8.2}
\end{equation}
The local inverse temperature $\beta(q) = \beta_{\delta}$ for $q
\in \Lambda_{\delta}$ and $\beta(q)$ interpolates smoothly otherwise.
For each choice of $\beta(q)$ we obtain a distinct action functional
$W(t)$. By local
conservation of energy, these differ only through boundary terms. For
later purposes it is convenient to choose a step interpolation as
$\beta(q) = \beta_{-}$ for $q_{1} < 0$ and $\beta(q) = \beta_{+}$ for
$q_{1} \ge 0$. Then the action functional is
\begin{equation}
        W(t) = (\beta_{+}-\beta_{-}) \int^t_{0}ds J_{0}(s)
        \label{8.3}
\end{equation}
with $J_{0}$ the energy current through the plane $\{q_{1}=0\}$,
\begin{eqnarray}
J_{0} & = & \sum^{N}_{j=1} {1 \over m} p_{j} \left\{ {1 \over 2m} p_{j}^2 +
{1 \over 2} \sum^{N}_{i \neq j =1} V(q_{i}-q_{j}) \right\} \delta
(q_{j1})\\
\nonumber
& + & {1 \over 2} \sum_{i,j=1}^{N} {1 \over 2m} [((p_{j} + p_{i}) 
\cdot F(q_{j} - q_{i}))
(q_{j} - q_{i}) \int^1_{0} d \lambda \delta (\lambda 
q_{j1}+(1-\lambda)q_{i1})\, ,
\label{8.4}
\end{eqnarray}
where $q_{j_1}$ is the $x$-component of the $j$th particle position vector.

As an advantageous numerical alternative deterministic thermostatting
reservoirs have been developed \cite{EM}. We mention here the proposal of
Galla\-vot-ti and Cohen \cite{GC} for thermal reservoirs. The equations of
motions are Newtonian in the bulk. At the boundary layers $\Lambda_{+}$ and
$\Lambda_{-}$ additional friction terms are added which are constructed in
such a way that the total kinetic energy of the particles in $\Lambda_{+}$,
resp. $\Lambda_{-}$, is kept constant. The particular boundary temperature
is then fixed through the initial conditions.  The equations of motion read
\begin{eqnarray}
        {d \over dt} q_{j} & = & {1 \over m}p_{j}\, , \nonumber \\
{d \over dt} p_{j} & = & F_{j} + F_{w}(q_{j}) + \sum_{\delta= \pm}
\chi_{\delta} 
        (q_{j}) \alpha_{\delta}p_{j}\, .
        \label{8.5}
\end{eqnarray}
The `friction' coefficients $\alpha_{+}, \alpha_{-}$ are given by
\begin{eqnarray}
 \alpha_{\delta} & = & \left[ \sum^{N}_{j=1} \chi_{\delta}(q_{j}) p_{j}^2 \right]
        ^{-1}
\sum^{N}_{j=1} \left\{ {1 \over m} p_{j} \cdot \nabla \chi_{\delta}(q_{j})
 {1 \over 2m} p_{j}^2 \right. \nonumber \\
 &&\left.  + {1 \over m} \chi_{\delta}(q_{j}) (F_{j}+F_{w}) \cdot p_{j} 
 \right\}\, ,\quad \delta = \pm 1 \, .
\label{8.6}
\end{eqnarray}

For thermostatted systems the action functional is the phase space
contraction integrated along a trajectory of the dynamical system,
which in our case equals
\begin{equation}
        3 N \sigma = 3N_{+}\alpha_{+} + 3N_{-}\alpha_{-} + {\cal O}(1)
        \label{8.7}
\end{equation}
 with $N_{\pm}$ the number of particles in $\Lambda_{\pm}$.  The
fluctuation theorem holds provided the dynamics is sufficiently hyperbolic
for the SNS to be described by an SRB measure.  This is not expected to be
true in general.  It is then an assumption, embodied in the chaotic
hypothesis \cite{G1}, that the fluctuation theorem remain valid for
realistic physical systems.  The numerical simulation in \cite{P} gives
strong support for the validity of the fluctuation theorem in one such
model. There the two heat reservoirs are linked through an anharmonic
chain, rather than a fluid. Further numerical support comes from a study of
shear flow for hard disks in a box \cite{CL} where momentum but no energy
is transferred to the system at opposite boundaries.  The fluctuation
theorem is well verified \cite{BCL}, although it seems to be difficult to
get beyond the quadratic approximation.

The action functional defined with the phase space contraction (7.7)
appears to be rather different than the one of (7.3). However the
conservation law can be used again to transform a boundary flux to an
interior flux.
Let $H_{1}$ be the total energy of particles in the left half
box $[-a-\ell ,0] \times [0,\ell]^2$. We have
\begin{eqnarray}
{d \over dt}H_{1}(t) & = & - J_{0}(t) - \left[ \sum^N_{j=1}
\chi_{-}(q_{j}) p_{j}^2 \right]
\alpha_{-} \nonumber\\
 & = & -J_{0}(t) - {1 \over \beta_{-}} 3N_{-} \alpha_{-}\, ,
        \label{8.8}
\end{eqnarray}
since $\sum^{N}_{j=1} \chi_{-} (q_{j}) p^2_{j}$ is a constant of motion and
initially fixed to $3N_{-}/\beta_{-}$.
We conclude that
\begin{equation}
\int^t_{0} ds 3N \sigma(s) = - H(t) + H(0) + (\beta_{+}-\beta_{-}) \int
        ^t_{0} ds J_{0}(s)\, .
        \label{8.9}
\end{equation}
Up to a boundary term the phase space contraction is just the energy flux
across the plane $\{ q_{1}=0 \}$, the same quantity which has been obtained
for stochastic reservoirs, compare with (4.7).

It is tempting to assume that for stochastic and thermostatted boundaries,
the rate functions of the fluctuation theorem are identical. This requires
however a very strong form of equivalence of ensembles, in the sense that
in the interior of the system even the large deviations of the current are
in the steady state independent of the mechanism by which the boundaries
are cooled/heated.  This appears to be the case for the model studied in
\cite{P}.  On the other hand recent simulations \cite{BONL} suggest that this
equivalence fails for the model described in \cite{CL}--\cite{BCL}.

\section{Green-Kubo formula, Onsager reciprocity}
\setcounter{equation}{0}

As observed in \cite{G1} the fluctuation theorem yields the Green-Kubo
formula at equilibrium. We give an alternate derivation in the spirit of
\cite{JK}. For the sake of concreteness we consider a $k$-species lattice
gas on a $d$-dimensional lattice. The lattice gas is driven by an external
field, cf. Section 3, and satisfies local detailed balance. For each
species the driving field is a $d$-component vector. It is convenient to 
regard the driving field as the $m$-component vector $\vec{F} =
(F_{1},\ldots,F_{m})$, $m = d \times k$. There are $m$ different types of
jump, labeled by the species and lattice directions. Therefore the
generator of the lattice gas is the sum
\begin{equation}
L = \sum_{j=1}^{m}L^{(j)}\, ,
\label{9.1}	
\end{equation}
where each $L^{(j)}$ satisfies the conditions of Section 2.3,
compare with (2.23). To each $L^{(j)}$ there is associated the current
$J_{j}(t)$. 

We define
\begin{equation}
e(\vec{F};\lambda_{1},\ldots,\lambda_{m}) =  
\lim_{t \rightarrow \infty} - {1 \over t} \log \langle \exp[- \beta
\sum_{j=1}^{m}\int_{0}^{t}  \lambda_{j}J_{j}(s)ds] \rangle_{\vec{F}} 
\, ,
\label{9.2}
\end{equation}
where we have, in the action, absorbed $F_j$ into the $\lambda_j$'s and
indicated explicitly the dependence of the stationary state and the dynamics on
$\vec{F}$.  The fluctuation theorem now takes the form
\begin{equation}
e(\vec{F}*;\lambda_{1},\ldots,\lambda_{m}) =  
e(\vec{F}*;F_{1} - \lambda_{1},\ldots,F_{m} - \lambda_{m})\, .
\label{9.3}
\end{equation}

We have  $\langle J_{j}(t)\rangle_{0} = 0$, since 
$\vec{F} = 0$ corresponds to equilibrium. Differentiating (8.2) 
at $\lambda =0$, we obtain 
\begin{equation}
        {\partial \over \partial \lambda_{j}} e(\vec{F}*;0) = 
		\beta \langle J_{j}(t)\rangle_{\vec{F}}\, ,
        \label{9.4}
\end{equation}
the average $j$-th current.

Now the linear response in the average current to a small driving field is
given by
\begin{equation}
{\partial \over {\partial F_{j}}}{1 \over \beta} 
{\partial \over \partial \lambda_{i}}
e(0;0) = \sigma_{ij}\, .
\label{9.5}
\end{equation}
Differentiating (8.3) with respect to $\vec{F}$ and $\vec{\lambda}$ 
at $\vec{F} = \vec{\lambda} = 0$
 we conclude that
\begin{equation}
\beta \sigma_{ij} = - \beta \sigma_{ij} +  \beta^2
\hat{\sigma}_{ij}
\label{9.6}
\end{equation}
and therefore
\begin{equation}
\sigma_{ij} = {1 \over 2}\beta \hat{\sigma}_{ij}\, ,
\label{9.7}
\end{equation}
where
\begin{equation}
\hat \sigma_{ij} = -{1 \over \beta^2} {\partial \over {\partial \lambda_{i}}}
{\partial \over {\partial \lambda_{j}}}
 e(0;0) =
        \int^\infty_{-\infty} dt \langle 
        J_{i}(t)J_{j}(0)\rangle_{0}\, .
        \label{9.8}
\end{equation}
This is the standard Einstein-Green-Kubo relation between
linear response in the current
and the time--integrated current--current correlation, which satisfies 
the Onsager relation $\hat \sigma_{ij} = \hat
\sigma_{ji}$.

We remark that the derivation of (8.7) differs considerably from the
standard computation, we refer to \cite{S} for an exposition of the latter.
There one differentiates, at finite volume, the steady state current
$\langle j_{\vec{F}}\rangle_{\vec{F}}$ at $\vec{F}=0$, which by definition
equals $\sigma_{ij}$.  This gives two terms: one from differentiating the
current function $j_{\vec{F}}$ and one from differentiating the steady
state $\langle\cdot\rangle_{\vec{F}}$.  Their sum is then (8.8), because
$\langle J_{i}(t)J_{j}(0)\rangle_{0}$ has a $\delta$-peak at $t = 0$, whose
weight is the first term, and a smooth piece, whose time integral equals
the second term.

\section{Conclusions}
\setcounter{equation}{0}

Within the framework of stochastic dynamics the fluctuation theorem holds
in great generality.  For finite state spaces, like stochastic lattice
gases, we have given a proof.  For the nondegenerate diffusion processes of
Section 5 we could provide sufficient conditions on the driving forces and
the diffusion matrix which ensure the validity of (5.8).  The situation is
more delicate for degenerate diffusion processes, in particular for
deterministic bulk systems driven by stochastic boundaries.  Since the
noise vanishes on large parts of the phase space, even the convergence to
the steady state is not obvious.  To our knowledge there are only a few
rigorous investigations of SNS of this type \cite{LLR,OL,GLP,GI,EPR}.  In
\cite{GLP,GI} classical point particles are considered. At a collision with
the wall they are reflected with a Maxwellian distribution at the local
temperature of the wall. Existence and uniqueness of the invariant measure
is established provided the forces are repulsive and have a range of the
size of the box. This latter property prevents pockets in phase space which
never see the wall. In \cite{LLR} explicit SNS measures are obtained for a
harmonic chain coupled to stochastic reservoirs specified by Langevin
forces at the ends of the chain.  While in \cite{EPR} an anharmonic chain
is coupled to a free field leading to similar boundary conditions.  For
this system existence and uniqueness of the invariant measure is
established in under a H\"{o}rmander condition.  We expect that for these
systems the GC fluctuation theorem can be proved.  Thermostatted boundaries
are beyond our mathematical abilities at the present time.

Viewed from the physics point of view the fluctuation theorem is a
consequence of time-reversal.  For a given stationary stochastic process
one considers the density of the path measure of the time-reversed process
relative to the original one.  The logarithm of this density is the action
functional which satisfies the fluctuation theorem.  At this level of
generality it is somewhat unexpected that in concrete models, which satisfy
the condition of local detailed balance, the action functional has a direct
physical interpretation.  By separating the rate of change of Gibbs entropy
into a production and flow term we identify the observable whose average
defines the entropy flow.  The action functional is then this observable
integrated along a stochastic trajectory.  Generically it is the current,
multiplied by the driving force, corresponding to the conserved field.  The
GC fluctuation theorem is then a symmetry property for the large deviations
in the current.  Such large deviations are not readily observed in physical
systems, because they refer to exponentially small probabilities. Also in
numerical experiments special efforts are needed \cite{ECM}, \cite{BCL},
\cite{BGGi}. Even worse, the action functional is extensive and the
probabilities are exponentially small also in $N$.  *Thus it would be
desirable to have a local quantity satisfying (at least approximately) the
fluctuation theorem but this does not appear to be the case for the systems
investigated numerically thus far \cite{BCL}, \cite{BONL}.

There is one prediction of the fluctuation theorem which is worthwhile to
emphasize.  We use the convention of the previous section where the
symmetry reads $e(\lambda) = e(F-\lambda)$, i.e.\ $e$ is even relative to
$\lambda = F/2$, and we may expand as $e(\lambda) = e_0 + {1 \over 2} e_2
\lambda(F - \lambda) + \ldots$. Assuming that in the interval $-\delta <
\lambda < F + \delta$ with some small $\delta > 0$ the quadratic
approximation is valid, then by differentiating at $\lambda = 0$ we obtain
$\langle j \rangle_F = e_2 F/2$, where $e_2 =  \int^\infty_{-\infty}
dt(\langle J(t) J(0) \rangle_F - 
\langle J(t) \rangle_F\langle J(0) \rangle_F)$ will generally
depend on $F$.  The nonlinear response in the current is then  linked to the
time-integral over the current-current correlation.  The knowledge of
$e(\lambda)$ thus provides information about the range of
validity of the Einstein relation and linear response theory.

\begin{appendix}
\section{Appendix: Large deviation function for the asymmetric exclusion
process} 
\setcounter{equation}{0}

To determine the large deviation function in any of the models we have
discussed is like computing a free energy in equilibrium statistical
mechanics.  Therefore it is not surprising that even in the stochastic
framework explicit results are available only for a few models.  We discuss
here the large deviation function for the asymmetric simple exclusion
process in one dimension where particles random walk with an average drift
and interact through the constraint that there can be at most one particle
per lattice site.  As a warm up we consider this model without constraint,
i.e.\ independent biased random walks on a lattice.  It would be of
interest to have also an example for the boundary driven lattice gases of
Section 4.  Besides independent particles, the simplest model is symmetric
exclusion in the bulk and boundary reservoirs as discussed in Section 4.
Its large deviaiton function, for $\tilde W_x(t)$, satisfies then $e(\lambda)
= e(\beta(\gamma_1 - \gamma_\ell) - \lambda)$, cf.\ (4.7).  Thus in the
limit $\gamma_1 \to \infty$, i.e.\ at the left boundary particles are only
injected, the GC fluctuation theorem predicts $\lim_{t \to \infty} (-1/t)
\log {\rm Prob}(\{\tilde W_x < at\}) = 0$ for $a < 0$.

We first consider $N$ independent particles, i.e.\ no exclusion, on a ring
of size $\ell$ driven by a uniform force of strength $F$.  Since particles
are independent, it suffices to study the one-particle problem.  For $N$
particles we merely have to multiply by $N$ at the end. It is convenient to
set $\beta F/2 = E$.  The generator for the dynamics reads
\begin{equation}
 L_{E} f(x) = (e^{E}/2 \cosh E)f(x+1) + (e^{-E}/2 \cosh E) f(x-1) - f(x)\,
  .  \label{5.1} \end{equation} The elementary jump time is normalized such
  that for $E \rightarrow \infty$ the jump rate to the right is one. $W(t)$
  is now the total number of signed jumps up to time $t$. To compute the
  generating function $\langle e^{-\lambda W(t)}\rangle$ for large $t$ we
  need the maximal eigenvalue of
\begin{equation}
L_{E,\lambda} f(x) = (e^{E -\lambda}/2 \cosh E) f(x+1) +
(e^{-E + \lambda}/2 \cosh E) f(x-1) - f(x)\, .
        \label{5.2}
\end{equation}
Clearly $f=1$ solves $L_{F,\lambda} f = -e(\lambda) f$ with
\begin{equation}
 e(\lambda) = 1 - (\cosh (E - \lambda) /  \cosh E)\, .
        \label{5.3}
\end{equation}
Its Legendre transform is given by
\begin{equation}
 \hat{e}(w) = - Ew +1 + w {\rm arg sinh}(w \cosh E) - \sqrt{w^{2} + (\cosh
 E)^{-2}}\, .
        \label{5.4}
\end{equation}
$\hat{e}$ vanishes at the average current $j = \tanh E$.
On general grounds we know that $e(\lambda)$ is convex
down, $e(0) = 0, e'(0) =-j$, the average current, $e(\lambda) =
e(2E-\lambda)$ and that $\hat{e}(w)$ is convex up, $\hat{e}(w) \ge 0 \ , \
\hat{e}(w_{0}) = 0$ for $w_{0}=j={1 \over t} \langle W(t)\rangle$, $\hat{e}
(w) - \hat{e}(-w) = -2Ew$. 

Note that in
the limit $E \rightarrow \infty$, $\hat{e}$ degenerates to
$\hat{e}(w) = 1 - w + w \log w$ for $w \geq 0$ and $\hat{e}(w) =
\infty$ for $w < 0$, which merely reflects that left jumps have
probability zero for $E = \infty$.

We expect more structure for particles interacting through 
hard core
exclusion. This corresponds to $H(\eta) = 0$ in the notation of Section
3.  The jump rates with a uniform driving field are then
\begin{equation}
        c_{xx+1}^E (\eta) = (e^E / 2 \cosh E) \eta_{x}(1-\eta_{x+1})+
        (e^{-E}/2\cosh E)(1- \eta
        _{x}) \eta_{x+1}\, .
        \label{5.5}
\end{equation}
We normalized the rates such that for $E \rightarrow \infty$ a
particle jumps with
unit rate to the right. (A.5) are the jump rates of the asymmetric
simple exclusion process (ASEP).

The action functional $W(t)$ is the total current integrated over
the time span $t$. Expectations, $\langle\cdot\rangle$, are taken
in the stationary process at a given number of particles $N$,
$1 \le N \le \ell - 1$. For the ASEP this means that at time $t = 0$
all allowed
configurations have equal weight. Then
\begin{equation}
\lim_{t \rightarrow \infty} - {1 \over t} \log \langle e^{-\lambda W(t)}
\rangle = e_{E}
        (\lambda, \ell, N)
        \label{5.6}
\end{equation}
and $-e_{E}$ is the maximal eigenvalue of $L_{\lambda}$ with
\begin{eqnarray}
L_{\lambda} f(\eta) = (2 \cosh E)^{-1} \sum^\ell_{x=1} \left\{ (e^{E-\lambda}
\eta_{x} (1- \eta_{x+1}) + e^{-(E - \lambda)} (1 - \eta_{x})
\eta_{x+1}) \right. \nonumber \\
\left.  f(\eta^{xx+1}) - (e^E \eta_{x}(1-\eta_{x+1}) + e^{-E}
(1-\eta_{x})\eta_{x+1}) f(\eta) \right\}\, .\quad
        \label{5.7}
\end{eqnarray}
The fluctuation theorem reads
\begin{equation}
e_{E}(\lambda) = e_{E}(2E - \lambda)\, ,
        \label{5.8}
\end{equation}
which can be seen also directly from (A.7).

$L_{\lambda}$ is a linear operator with a structure which is similar to that
of the hamiltonian  for a quantum spin chain. To make contact with
that model  we rewrite
$L_{\lambda}$ in terms of the Pauli spin ${1 \over 2}$ matrices $
\vec{\sigma}_{x}=
(\sigma^1_{x}, \sigma^2_{x}, \sigma^3_{x})$ associated to every
lattice site $x$. We
  identify $\eta_{x} = 1$ with spin up, $\sigma^3_x = 1$, and $\eta_{x}
= 0$ with spin down, $\sigma^3_x = -1$. Then
  \begin{eqnarray}
    -L_{\lambda} = H_{\lambda} = \sum^\ell_{x=1}
    \left\{
    (1/4) (1 - \sigma_{x}^3
    \sigma_{x+1}^3) - (e^{E - \lambda}/2 \cosh E) \sigma_{x}^+
    \sigma_{x+1}^- \right.
    \nonumber \\
  \left. - (e^{- E + \lambda}/2 \cosh E) \sigma^-_{x} \sigma^+_{x+1}
   \right\}\,  ,
        \label{5.9}
\end{eqnarray}
where $\sigma_{x}^\pm = {1 \over 2} (\sigma^1_{x} \pm i \sigma_{x}^2)$ are
the spin
raising and lowering operators. We note that $H^\ast_{\lambda} \neq
H_{\lambda}$, 
unless $E = \lambda$, and $H_{\lambda}$ does not have directly a
quantum mechanical interpretation, except through analytic
continuation in $\lambda$.

To determine the ground state energy of $H_{\lambda}$, the only technique
available is the Bethe ansatz. Fortunately, $H_{\lambda}$ is covered by the
famous paper of Sutherland and Yang \cite{SYY}. However, the analysis of
the nonlinear Bethe equations still require a considerable effort, which in
our section of parameter space has been carried out only recently. We use
the results by D.Kim \cite{K}. He is in fact interested in the energy gap,
but also gives $e_{E} (\lambda, \ell, N)$ to leading order in $N$ in the
limit $\ell \rightarrow \infty,$ $N \rightarrow \infty $, $\rho = N/\ell$
fixed.  We expect that also order 1 terms could be obtained, but such an
analysis is certainly beyond the scope of our paper. In the limiting case
$E \rightarrow \infty$, the Bethe equations simplify. This has been
exploited recently by Derrida et al. \cite{DL,DA}. They determine
$e_{\infty}(\lambda, \ell,N)$ fairly explicitly and discuss the order 1
corrections to the infinite volume limit.

To leading order in $N$, $N/\ell = \rho$, we have
\begin{equation}
        e_{E}(\lambda, \ell, N) \cong \ell \bar{e}_{E}(\lambda, 
        \rho)\, ,
        \label{5.10}
\end{equation}
i.e. $e_{E}$ is extensive. $\bar{e}_{E}$ has a flat piece,
$\bar{e}_{E}(\lambda, \rho) = 0$ for $0 \le \lambda \le 2 E$.
At $\lambda = 0, \bar{e}_{E}$ has the slope
$\rho(1-\rho) \tanh E$ and the asymptotics \cite{K}
\begin{eqnarray}
\bar{e}_{E}(\lambda, \rho) & = &\rho(1-\rho)(\tanh E) \lambda \\
& - & {1 \over 20}
\left( {3 \pi \over 2} \right)^{2/3} (4 \rho(1 - \rho))^{4/3} (\tanh E)
        |\lambda|^{5/3}(1 + {\cal O}(|\lambda|^{2/3})) \nonumber
        \label{5.11}
\end{eqnarray}
for $\lambda < 0$. By symmetry $\bar{e}_{E}$ has the same asymptotics
at $\lambda = 2E$. For $E \rightarrow \infty$, the right zero $e_{E}
(\lambda, \ell,
N)$ moves to $\infty$ and the flat piece of $e_{E}(\lambda, \ell,
N)$ never levels off, in agreement with \cite{DL,DA}.

$\bar{e}'(0)$ is just the average current which is $\rho(1 - \rho)
\tanh E$. $\bar{e}''(0)$ is formally the integral over the total
current-current
correlation and therefore the mobility of the lattice gas \cite{S}.
In our case $\bar{e}''(0) = \infty$, which reflects the fact
that in the ASEP density
fluctuations propagate superdiffusively \cite{BKS}.

At $E =0$ (A.6) gives the large deviations in the total current at
equilibrium, i.e. for the symmetric simple exclusion process. The
fluctuation theorem reduces then to $e_{0,\ell,N}
(\lambda) = e_{0,\ell,N}(- \lambda)$, which also follows from time
reversibility. 
$\bar{e}_{0}(\lambda,
\rho)$ has no longer a flat piece. For small $\lambda, \bar{e}_{0} 
(\lambda, \rho)
= - {1 \over 2}
\lambda^2 \rho(1 - \rho)$.  Thus the mobility $\sigma(\rho) = \rho
(1 - \rho)$ and the diffusion constant, which is $\sigma$
divided by the compressibility,  $D(\rho) = 1$, independent of the density
in agreement with the known
bulk diffusion of the symmetric simple exclusion process. The more precise
asymptotics \cite{K} is given by
\begin{equation}
\bar{e}_{0} (\lambda, \rho) = - {1 \over 2} \rho(1 - \rho) \lambda^2 -
{1 \over 20} 2^{1/3} (2 \pi)^{2/3}(\rho(1-\rho))^{4/3}|\lambda|^{8/3}
(1+ {\cal O}(|\lambda|^{4/3}).
                \label{5.12}
\end{equation}
On a formal level the ``Burnett coefficients'' correspond to higher
derivatives of $\bar{e}_{E}(\lambda)$ at $\lambda = 0$ \cite{vB}. (A.12)
shows that for the symmetric simple exclusion process the Burnett
coefficients are infinite.

Just as free energies, the large deviation function may become singular in
the infinite volume limit and this happens already for the first nontrivial
example. To understand the origin of this behavior, it is necessary to
analyze those configurations which give the main contribution to $\langle
e^{-\lambda W(t)}\rangle$.  If $\lambda \rightarrow \infty$ only left jumps
are permitted and similarly for $\lambda \rightarrow - \infty$ only right
jumps are permitted, essentially independent of the driving field $E$, $E >
0$ for the sake of discussion. However, in the intermediate regime $0 \le
\lambda \le 2E$, the particles have two opposing instructions. Should they
follow $E$ or $\lambda$? In fact, neither. As can be inferrred from the
maximal eigenvector, they just stick and form one big cluster
\cite{YY}. The clustering is seen most easily at $\lambda = E$, where
$H_{\lambda}$ is the ferromagnetic, anisotropic Heisenberg model and the
maximal eigenvector its ground state. At our parameters, $H_{\lambda}$ is
in the ferromagnetic phase. Since the magnetization is fixed, typical
configurations condense into one large cluster. In our units the
corresponding ground state energy is of order 1 independent of $\ell$.

It is instructive to reconsider our result from the point of view of the
probability distribution of $W(t)/t$, which is obtained from the
Legendre transform of $\bar{e}_{E}(\lambda)$. To make it well defined,
we note that, for large $N$,  $e_{E}(\lambda)$ smoothly interpolates between
$e_{E}(0) = 0 = e_{E}(2E)$, such that $e'_{E}(0)$ $= N j(\rho), e'_{E}(2E)
= - N j(\rho), j(\rho) = \rho(1-\rho) \tanh E$, and $e_{E}(E)$ =
${\cal O}(1)$.
Thereby we obtain, for $E \ge 0$,
\begin{equation}
\hat{\overline{e}}_{E}(w, \rho) =
\left\{ \begin{array}{l@{\quad \textrm{for} \quad}l}
     0 & 0 \le w \le j(\rho) \, , \\
     -2E w & - j(\rho) \le w \le 0 \, .
     \end{array} \right.
        \label{5.13}
\end{equation}
For $w$ slightly larger than $j(\rho)$ we have $
\hat{\overline{e}}_{E}(w, \rho) \simeq (w - j(\rho))^{5/2}$  and
$\hat{\overline{e}}_{E}(w, \rho)$ $ \simeq w\log w$ in the limit of large $w$.
 For  $w < - j(\rho)$ the values of $\hat{\overline{e}}_{E}$
are fixed by the fluctuation theorem.

To understand the $N$-dependence of $\hat{e}_{E}(w,\ell,N)/\ell$, we note
that for $W(t)/$ $Nt$ to have a value larger than $j(\rho)$ we have to
speed up all $N$ particles. For it to have a value smaller than $j(\rho)$
it suffices to slow down a single particle, since the other particles pile
up behind. Typical configurations consist of essentially a single
cluster. The large deviation rate is ${\cal O}(1/N)$ which on our scale
corresponds to $\hat{\overline{e}}_{E} = 0$. Such a mechanism works only
for $(W(t)/Nt) \ge 0$. To have a negative total current, again all $N$
particles are forced to move, now opposite to $E$. The linear decrease of
$\hat{e}_E(w<N,\ell)/\ell$ with corrections of ${\cal O}(1/N)$ for
$-j(\rho) \leq w \leq 0$ follows from the fluctuation theorem.  In the
limit $E \rightarrow \infty$, $\hat{e}_{E}(w,\ell,N) = \infty$ for $w < 0$.
This just reflects the fact that the underlying process does not allow for
histories with negative currents.

{}From our discussion of $\hat{\overline{e}}$ we conclude that the
non-smooth behavior of $\bar{e}(\lambda)$ for large $N$ comes from blocking
through slow particles, which suggests that, if we go to higher spatial
dimension or soften the hard core, a smooth large deviation function may be
recovered. But this remains to be seen. The symmetric case, $E = 0$, warns
against too fast a conjecture.\bigskip\\
\end{appendix}
\noindent
{\large \textbf{Acknowledgments.}} We are grateful to F. Bonetto,
G. Gallavotti, C. Maes, and D. Ruelle for instructive discussions.  The
research was supported by NSF Grant NSF-DMR 95-23266 and AFOSR Grant
F49620-98-1-0207.  We also acknowledge DIMACS and its supporting agencies,
the NSF under contract STC--91--19999 and the NJ Commission on Science and
Technology.

\end{document}